\documentclass[12pt]{iopart}
\usepackage{iopams}
\usepackage{amsfonts}
\usepackage{amsthm}
\usepackage{amstext}
\usepackage{amssymb}
\usepackage{amsbsy}
\usepackage{amscd}
\usepackage{eufrak}
\usepackage{latexsym}
\usepackage[dvips]{graphics}
\usepackage[pdf]{graphicx}
\usepackage{a4wide}
\usepackage{geometry}
\usepackage{epsfig}
\usepackage{fancyhdr}
\usepackage{color} 
\usepackage[colorlinks]{hyperref}
\usepackage{pstricks}
\usepackage{indentfirst}
\usepackage{subfigure}
\usepackage{psfrag}
\usepackage{ifthen} 
\usepackage{eso-pic}
\usepackage[all]{xy}
\input amssym.def
\input amssym.tex
\eqnobysec
\begin{document}
\title{Integrability of a conducting elastic rod in a magnetic field}
\author{D.~Sinden and G.~H.~M.~van~der~Heijden}
\address{Centre for Nonlinear Dynamics, University College London, Chadwick Building, Gower Street, London WC1E 6BT, UK} 
\eads{\mailto{d.sinden@ucl.ac.uk}}
\begin{abstract}
We consider the equilibrium equations for a conducting elastic rod placed in a uniform magnetic field, motivated by the problem of electrodynamic space tethers. When expressed in body coordinates the equations are found to sit in a hierarchy of non-canonical Hamiltonian systems involving an increasing number of vector fields. These systems, which include the classical Euler and Kirchhoff rods, are shown to be completely integrable in the case of a transversely isotropic rod; they are in fact generated by a Lax pair. For the magnetic rod this gives a physical interpretation to a previously proposed abstract nine-dimensional integrable system. We use the conserved quantities to reduce the equations to a four-dimensional canonical Hamiltonian system, allowing the geometry of the phase space to be investigated through Poincar\'e sections. In the special case where the force in the rod is aligned with the magnetic field the system turns out to be superintegrable, meaning that the phase space breaks down completely into periodic orbits, corresponding to straight twisted rods.
\end{abstract}
\newtheorem{thm}{Theorem}[section]
\newtheorem{conj}[thm]{Conjecture}
\newtheorem{cor}[thm]{Corollary}
\newtheorem{lem}[thm]{Lemma}
\newtheorem{prop}[thm]{Proposition}
\newtheorem{rem}[thm]{Remark}
\newtheorem{defin}[thm]{Definition}
\newtheorem{cla}[thm]{Claim}
\newtheorem{example}[thm]{Example}
\section{Introduction} \label{sec:intro}
The problem of an elastic conducting wire in a magnetic field is a classical
one in magnetoelasticity. For instance, it is well-known that a straight
current-carrying wire held between pole faces of a magnet buckles into a
coiled configuration at a critical current
\cite[Section 10.4.3]{Woodson68}. A rigorous bifurcation analysis of
this buckling problem (for a uniform magnetic field directed parallel to the
undeformed wire) was developed in a series of papers by Wolfe. He first
considered a nonlinearly-elastic string model for the wire, i.e., a perfectly
flexible elastic line, and found that an infinite number of solution branches
bifurcate from the trivial straight solution \cite{Wolfe83,Wolfe90b}, much like
in the Euler elastica under compressive load. The equations can in fact be
solved exactly and the non-trivial solutions are exact helices. In subsequent
work Wolfe modelled the wire as a rod \cite{Wolfe88,Seidman88,Wolfe96}. In addition
to extension a rod can undergo flexure, torsion and shear, and for the case of
welded boundary conditions it was found that in certain cases bifurcation occurs,
again with an infinity of non-trivial equilibrium states.
In all these studies the wire is assumed to be non-magnetic. Magnetoelastic
instability of a current-carrying rod due to its own magnetic field has also
been considered \cite{Moon84,Chattopadhyay79,Smits88}.

The problem of a conducting rod in a magnetic field has recently attracted
interest as a model for electrodynamic space tethers, i.e., conducting cables
that exploit the earth's magnetic field to generate thrust and drag (Lorentz)
forces for manoeuvring \cite{Valverde03,Heijden05}.

Here we consider the hitherto unexplored Hamiltonian structure and
integrability of the equilibrium equations for a rod in an external magnetic field.
Like Wolfe we assume the magnetic field to be uniform and we shall concentrate on
the usual case of a rod that is uniform, intrinsically straight, inextensible and
unshearable. For this case the equilibrium equations formulated in body coordinates
are found to sit in a hierarchy of rod models, described by an increasing number of
three-dimensional field vectors, with an interesting non-canonical Hamiltonian structure.
The first two members of this hierarchy are the classical single-field force-free (Euler)
rod and the two-field Kirchhoff rod (the two fields being the force and moment in the rod).
Both these rods have well-known dynamical analogues in, respectively, the swinging
planar pendulum and the spinning top \cite{Love44,Kehrbaum97b}. The
`magnetic rod', the next level in the hierarchy, does not seem to have a
dynamical analogue. However, it does give a physical realisation to (a special
case of) the `twisted top' introduced in \cite{Thiffeault01}. The equations
of this top were introduced as an abstract mathematical construction obtained
by extending the Lie-Poisson bracket of the heavy top by a cocycle
(Leibniz extension). The resulting three-field bracket is the simplest
non-semidirect extension of the heavy-top bracket \cite{Thiffeault00} and the
authors in \cite{Thiffeault01} were interested in how this extension affects
the dynamics of the system. We show that the `twisted top' is in fact a magnetic rod.

By using the Casimir invariants for the magnetic rod we perform an explicit
reduction to a six-dimensional canonical Hamiltonian system with a two first integrals in addition
to the Hamiltonian. This reduction is valid for all but an invariant subspace of
the phase space where the magnetic field is aligned with the force in the rod.

The magnetic rod can be regarded as a perturbation of the Kirchhoff rod and is
found to retain the latter's property that the equations are completely integrable
if the rod is transversely isotropic, i.e., if the principal bending stiffnesses
are equal (the dynamical analogue is the symmetric Lagrange top). In fact, the
subhierarchy of isotropic rods is generated by a Lax pair. On the alignment
subspace the isotropic rod is found to be superintegrable and the equations can be
solved directly by algebraic means. All solutions correspond to twisted straight rods.

The next (four-field) level of the hierarchy also has physical meaning:
it describes a rod in a special type of non-uniform magnetic field generated
by what might be called a uniform hypermagnetic field. Alignment of these two fields again
defines a superintegrable case with a breakdown of the phase space into periodic orbits
corresponding to twisted straight rods. The mathematical structure of the hierarchy
extends beyond this four-field model but the physical meaning becomes less clear.

The outline of the paper is as follows. In Section 2 we set up the
geometrically exact Cosserat theory used for modelling a rod. The hierarchy,
with increasing number of fields, is presented in Section 3. We briefly review the
classical cases in the present context and introduce the magnetic rod in Subsection 3.3.
We show that the isotropic case is integrable in the sense of Arnold-Liouville.
In Section 4 we perform the reduction of the magnetic rod equations to a six-dimensional
canonical Hamiltonian system. We also identify the superintegrable case. Section 5 presents
the hypermagnetic rod, while Section 6 gives the Lax pair for the subhierarchy of isotropic rods.

\section{Cosserat rods} \label{sec:model}

Cosserat theory for slender rods describes a rod as an arbitrary curve in space endowed with mechanical properties such as bending and torsional stiffnesses \cite{Antman95}. Forces and moments, whose balance equations will be introduced in the following sections, are assumed to be suitable averages over the rod's cross-section and to act at the centreline. The configuration of a rod is defined by a vector function $\boldsymbol{r}\left(s\right)$, describing the centreline of the rod, and an attached right-handed orthonormal triad of directors $\left\{ \boldsymbol{d}_{1}(s), \boldsymbol{d}_{2}(s), \boldsymbol{d}_{3}(s) \right\}$ describing the orientation of the cross-section. We assume the rod to be inextensible and unshearable. In mathematical terms this is expressed by the conditions $|\boldsymbol{r}'|=1$ and $\boldsymbol{r}'\cdot\boldsymbol{d}_1=\boldsymbol{r}'\cdot\boldsymbol{d}_2=0$, respectively, where $()'=\mbox{d}/\mbox{d}s$. Thus we can write $\boldsymbol{r}'=\boldsymbol{d}_3$ and the parameter $s$ measures arclength along the rod in any configuration. It will be convenient to express components of vectors with respect to the director (or body) frame; for any vector $\boldsymbol{v}$ the triple of components $v_i=\boldsymbol{v}\cdot\boldsymbol{d}_i$ ($i=1,2,3$) will be denoted by the sans-serif symbol $\mathsf{v}$.
\par
On introducing a right-handed orthonormal frame $\left\{ \boldsymbol{e}_{1}, \boldsymbol{e}_{2}, \boldsymbol{e}_{3} \right\}$ fixed in space we can write
\begin{eqnarray}
\boldsymbol{d}_{i} & = {R}\boldsymbol{e}_{i} \quad\quad (i = 1,2,3),
\label{eq:frame}
\end{eqnarray}
where $R$ is a rotation matrix. Differentiation gives
\begin{eqnarray}
\boldsymbol{d}^{\prime}_{i} & = R^{\prime}\boldsymbol{e}_i=R^{\prime}R^{-1}\boldsymbol{d}_i=:\boldsymbol{u} \times \boldsymbol{d}_{i},
\label{eq:directors}
\end{eqnarray}
where $\boldsymbol{u}$ is the vector of generalised strains. The body components $u_i=\boldsymbol{u}\cdot\boldsymbol{d}_{i}$ are the curvatures ($i=1,2$) and the twist ($i=3$) of the rod, for which, from \eref{eq:directors}, we can write
\begin{eqnarray}
u_{i}^{} & = \frac{1}{2} \varepsilon_{ijk}^{} \boldsymbol{d}^{\prime}_{j} \cdot \boldsymbol{d}_{k}^{},
\label{eq:curvatures}
\end{eqnarray}
where $\varepsilon_{ijk}$ is the standard Levi-Civita permutation symbol.
\par
Constitutive relations specify how the strains relate to the stresses on the rod. Here we assume the rod to be uniformly hyperelastic, i.e., we assume that there is a strain energy function $W(\mathsf{u})$ and that the stresses (body moments) $m_i$ can be obtained by partial differentiation:
\begin{eqnarray}
m_{i} &= \frac{\partial W}{\partial u_{i}}.\nonumber
\end{eqnarray}
We shall only consider the quadratic form
\begin{eqnarray}
{W} \left( {\mathsf{u}} \right) & = \frac{1}{2}K_1u_{1}^{2} + \frac{1}{2}K_2u_{2}^{2} + \frac{1}{2}K_3u_{3}^{2}
\label{eq:strain_energy}
\end{eqnarray}
for a uniform, intrinsically straight and prismatic rod, corresponding to the linear constitutive relations (Hooke's law)
\begin{eqnarray}
m_1=K_1u_1, \quad m_2=K_2u_2, \quad m_3=K_3u_3,
\label{eq:constit}
\end{eqnarray}
where $K_{1}$ and $K_{2}$ are the bending stiffnesses, about $\boldsymbol{d}_1$ and $\boldsymbol{d}_2$ respectively, and $K_3$ is the torsional stiffness, about the tangent $\boldsymbol{d}_3$. The rod is called transversely isotropic (or isotropic for short) if $K_1=K_2=:K$. For rods of homogeneous cross-section we have the relations: $K_1=EI_1$, $K_2=EI_2$, $K_3=GJ$, where $E$ is Young's modulus, $G$ the shear modulus, $I_1$ and $I_2$ the second moments of area of the cross-section about $\boldsymbol{d}_2$ and $\boldsymbol{d}_1$, respectively, and $J$ the polar second moment of area.

\section{A hierarchy of rod models}\label{sec:hier}
\subsection{The force-free rod -- Single-field model} \label{subsec:force_free}
The equilibrium equation for a force-free rod is
\begin{eqnarray}
\boldsymbol{m}^{\prime} & = \boldsymbol{0}.
\label{eq:fixed_frame}
\end{eqnarray}
In the director frame the equation can be written as a non-canonical Hamiltonian system
\begin{eqnarray}
\mathsf{m}^{\prime} & = J\left(\mathsf{m}\right) \nabla \mathcal{H},
\label{eq:noncan}
\end{eqnarray}
where the skew-symmetric structure matrix $J= J\left(\mathsf{m}\right)$ is given by
\begin{eqnarray}
J & = -J^{T} =
\left( 
\begin{array}{ccc}
0 & -m_{3} & m_{2} \\
m_{3} & 0 & -m_{1} \\
-m_{2} & m_{1} & 0 
\end{array} 
\right)
=:\hat{\mathsf{m}}
\label{eq:twist_structure}
\end{eqnarray}
and the Hamiltonian is
\begin{eqnarray}
\mathcal{H} & = \frac{1}{2} \mathsf{m} \cdot \mathsf{u},
\label{eq:twisted_hamiltonian}
\end{eqnarray}
with the $u_i$ as given in \eref{eq:constit}. For any two functions of $\mathsf{m}$ we can also introduce the Lie-Poisson bracket
\begin{eqnarray}
\left\{ f , g \right\}_{\left(\mathsf{m}\right)} & = - \underbrace{ \mathsf{m} \cdot \left( \nabla_{\mathsf{m}}f \times \nabla_{\mathsf{m}} g \right) }_{\mbox{twist}}, 
\label{eq:twisted_bracket}
\end{eqnarray} 
and write \eref{eq:noncan} as
\begin{eqnarray}
\mathsf{m}^{\prime} & = \left\{ \mathsf{m}, \mathcal{H} \right\}_{(\mathsf{m})} = \mathsf{m} \times \mathsf{u},
\label{eq:twisted}
\end{eqnarray}
which, on using \eref{eq:constit}, is also Euler's equation for the rotational motion of a free rigid body about its centre of mass.

For a general Hamiltonian system with Poisson bracket $\left\{\cdot,\cdot\right\}$ and Hamiltonian $\mathcal{H}$ a first integral is a function $I$ satisfying $\left\{I,\mathcal{H} \right\}=0$. Two integrals $I_1$ and $I_2$ are said to be in involution (or to commute) if $\left\{I_1,I_2\right\}=0$. In particular, if the bracket is non-canonical there exist conserved quantities, called Casimirs, that are in involution with any smooth function. Equivalently, $C$ is a Casimir if $\nabla C$ is in the null-space of the structure matrix $J$ \cite{Olver93}.  An $m$-dimensional non-canonical Hamiltonian system is completely integrable in the sense of Arnold-Liouville \cite{Arnold89} if it possesses $k$ Casimirs and $l$ first integrals (including the Hamiltonian), all in involution and with linearly independent gradients, such that
\begin{eqnarray}
m & = 2l + k.
\label{eq:count}
\end{eqnarray}
In particular, the solution of completely integrable systems can be reduced to quadrature.

The null-space of the structure matrix \eref{eq:twist_structure} is spanned by
\begin{eqnarray}
\nabla C_{1} & = \frac{1}{2} \left( m_{1}, m_{2}, m_{3} \right)^{T} \nonumber
\end{eqnarray}
and hence
\begin{eqnarray}
C_{1} & = \mathsf{m} \cdot \mathsf{m}
\label{eq:twisted_casimir}
\end{eqnarray}
is a Casimir. It describes the fact that the magnitude of the total moment is constant along the rod. Since $\mathcal{H}$ is an integral, it follows from \eref{eq:count} that \eref{eq:twisted} is completely integrable, a well-known fact from rigid-body dynamics. If $C_{1}=0$ then the system is degenerate.
\par
The force-free rod is not merely integrable, it is {\em superintegrable}, meaning that it has more independent integrals, $r$, than the number of degrees of freedom, $d$. Although independent, the integrals will not be in involution. In general, in a superintegrable system $2d-r$ conserved quantities define $r$-tori in $2d$-dimensional phase space. A system is {\em minimally superintegrable} if solutions exist on $(d-1)$-tori and {\em maximally superintegrable} if solutions exist on $1$-tori \cite{Fasso96}. The force-free rod has three degrees of freedom and there are four independent integrals: the Hamiltonian $\mathcal{H}$ and the three components of the moment $\boldsymbol{m}$ in the fixed frame (from \eref{eq:fixed_frame}). Hence, phase-space trajectories are restricted to invariant $2$-tori within the Liouville $3$-tori \cite{Hanssmann05} and the system is minimally superintegrable. In the isotropic case $\kappa^2=(m_1^2+m_2^2)/K^2$ and $\tau=m_3/K_3$ are constant and satisfy $K^2\kappa^2+K_3^2\tau^2=C_1$, and the rod configurations are twisted helices of curvature $\kappa$ and twist $\tau$ \cite{Nizette99}.
\par
In the special case where $K_{1} = K_{2} = K_{3} = K$ the equation in the body frame becomes $\mathsf{m}^{\prime}=\mathsf{0}$. This introduces a further, fifth, independent integral (we can choose the Hamiltonian, two components of the moment $\boldsymbol{m}$ in the fixed frame and two components of $\boldsymbol{m}$ in the body frame as independent integrals) and the phase space is foliated by $1$-tori. This case is maximally superintegrable. The corresponding rod configurations are resonant helices that have the material frame moving with the Frenet frame, with the twistless ring ($\tau=0$) and the twisted straight rod ($\kappa=0$) as extreme cases.
\subsection{The Kirchhoff rod -- Two-field model} \label{subsec:kirchhoff}
The equilibrium equations are
\begin{equation}
\boldsymbol{m}^{\prime} + \boldsymbol{r}^{\prime} \times \boldsymbol{n}=\boldsymbol{0}, \quad\quad \boldsymbol{n}^{\prime} = \boldsymbol{0},\
\label{eq:kirchhoff_external}
\end{equation}
where $\boldsymbol{m}$ and $\boldsymbol{n}$ are the moment and force in the rod. In the director frame the equations can be written in Hamiltonian form
\begin{eqnarray}
\left( 
\begin{array}{c}
\mathsf{m} \\
\mathsf{n} 
\end{array}
\right)^{\prime} & = 
J\left(\mathsf{m}, \mathsf{n}\right) \nabla \mathcal{H}\left( \mathsf{m}, \mathsf{n}\right),
\end{eqnarray}
where the structure matrix $J = J\left(\mathsf{m},\mathsf{n}\right)$ is given by
\begin{eqnarray}
J & = -J^{T} = 
\left( \begin{array}{cc} 
\hat{\mathsf{m}} & \hat{\mathsf{n}} \\
\hat{\mathsf{n}} & \mathsf{0}
\end{array}
\right)
\label{eq:kirchhoff_structure}
\end{eqnarray}
and the Hamiltonian is now
\begin{eqnarray}
\mathcal{H} & = \frac{1}{2} \mathsf{u} \cdot \mathsf{m} + \mathsf{d}_{3} \cdot \mathsf{n}, \quad \mbox{where} \quad \mathsf{d}_{3} = \left( 0,0,1 \right).
\label{eq:kirchhoff_hamiltonian}
\end{eqnarray}
Canonical and non-canonical Hamiltonian formulations of this problem have also been discussed in \cite{Kehrbaum97b}. Alternatively, we can write
\begin{eqnarray} \eqalign{
\mathsf{m}^{\prime} & = \left\{\mathsf{m}, \mathcal{H} \right\}_{(\mathsf{m},\mathsf{n})} = \mathsf{m} \times \mathsf{u} + \mathsf{n} \times \mathsf{d}_{3} \\
\mathsf{n}^{\prime} & = \left\{\mathsf{n}, \mathcal{H} \right\}_{(\mathsf{m},\mathsf{n})} = \mathsf{n} \times \mathsf{u},}\label{eq:kirchhoff}
\end{eqnarray}

in terms of the Lie-Poisson bracket on $\left( \mathsf{m}, \mathsf{n} \right)$ given by
\begin{eqnarray}
\left\{ f , g \right\}_{\left({\mathsf{m}},{\mathsf{n}}\right)} & = - \mathsf{m} \cdot \left( \nabla_{\mathsf{m}}f \times \nabla_{\mathsf{m}} g \right) - \underbrace{ \mathsf{n} \cdot \left( \nabla_{\mathsf{m}}f \times \nabla_{\mathsf{n}} g + \nabla_{\mathsf{n}}f \times \nabla_{\mathsf{m}} g\right) }_{\mbox{force}},
\label{eq:kirchhoff_bracket}
\end{eqnarray}
where an extra (semidirect) term has been added compared to \eref{eq:twisted_bracket}. The equations \eref{eq:kirchhoff} are also those for the  motion of a heavy top.
\par
The null-space of the structure matrix $J$ is two-dimensional and spanned by
\begin{equation}
\nabla C_{1} = \left(\begin{array}{c} \mathsf{n} \\ \mathsf{m} \end{array} \right) \quad \mbox{and} \quad \nabla C_{2} = \frac{1}{2} \left(\begin{array}{c} \mathsf{0} \\ \mathsf{n} \end{array} \right) \nonumber
\end{equation}
and hence the Casimirs are 
%
% \begin{subequations}
\begin{eqnarray}
C_{1} & = \mathsf{n} \cdot \mathsf{m} \label{eq:kirchhoff_casimir1}, \\
C_{2} & = \mathsf{n} \cdot \mathsf{n} \label{eq:kirchhoff_casimir2}.\
\end{eqnarray} 
\label{eq:kirchoff_casimirs}
% \end{subequations}
%
The Casimir $C_1$ describes the conservation of the moment about the force vector, while $C_2$ describes the conservation of the magnitude of force in the rod. If $C_2=0$ then the rank of the structure matrix changes and we recover the previous case of the force-free rod.
\par
Casimirs are conserved quantities that only depend on the structure of the balance equations; first integrals are dependent on the specific constitutive relations chosen. In addition to the Hamiltonian and the two Casimirs a first integral is required if the system is to be completely integrable. There are two cases, both well-documented:
\begin{description}
\item[The Lagrange case] has an integral given by
\begin{eqnarray}
I_{1} &= K\mathsf{m} \cdot \mathsf{d}_3 \quad\quad (\mbox{if}~~K_{1} = K_{2} = K), \nonumber
\end{eqnarray}
%
%where we have included the constant $K$ for reasons of consistency, which will become clear later.
Thus, if the two bending stiffnesses are equal then the twist $m_3$ is a conserved quantity.
\item[The Kovalevskaya case] has an integral given by
\begin{eqnarray}
\fl
I_{1} & = \left( K_{1}^{2}m_{1}^{2} - K_3^{2} m_{3}^{2} + n_{3} \right)^{2} 
+ \left( 2 K_{1}^{} K_3 m_{1}m_{3} - n_{1} \right)^{2} \quad\quad (\mbox{if}~~K_{1} = K_3 = 2K_{2}).
\label{eq:kov}
\end{eqnarray}  
The above condition on the bending stiffnesses renders the Kovalevskaya rod somewhat unphysical since it correspomds to a negative Poisson ratio. This ratio, defined as $\nu=E\slash{2G}-1$, is positive for ordinary materials. However, novel materials with negative effective Poisson ratio are now well known. For instance, experimental measurements of bending and torsional stiffnesses of DNA molecules have led to the generally accepted range $0.7<K/K_3<1.5$ \cite{Schlick95}.
\end{description}
There is another case which is not completely integrable on the entire phase space but is dependent on the value of Casimir $C_1$:
\begin{description}
\item[The Chaplygin-Goryachev case] requires that the initial conditions must satisfy
\begin{eqnarray}
\mathsf{m} \cdot \mathsf{n} & = 0 \nonumber;
\end{eqnarray}
then we have the integral
\begin{eqnarray}
I_{1} = K_{2}^{} m_{2} \left( K_{1}^{2} m_{1}^{2} + K_{2}^{2}m_{2}^{2} \right) - K_3 m_{3} n_{2} \quad\quad (\mbox{if}~~K_{1} = 4 K_{2} = K_3). \nonumber
\end{eqnarray}
\end{description}
It is simple to show that the integrals are in involution with respect to the bracket in \eref{eq:kirchhoff_bracket} and that $\mathcal{H}=h$, $C_{1}=\alpha$ and $I_{1}=\beta$ define three-tori on which configurations generically exist.
\subsection{The conducting rod in a uniform magnetic field -- Three-field model} \label{subsec:magnetic}
Now consider a rod placed in a uniform magnetic field $\bar{\boldsymbol{B}}$. The rod carries a uniform current $\boldsymbol{I}=I\boldsymbol{r}^{\prime}$ of strength $I$ along the centreline (we assume the rod to be sufficiently slender for eddy currents within the cross-section to be ignorable). The rod then experiences a Lorentz body force $\boldsymbol{F}_L=\boldsymbol{I}\times\bar{\boldsymbol{B}}=I\boldsymbol{r}^{\prime}\times\bar{\boldsymbol{B}}=I\boldsymbol{d}_3\times\bar{\boldsymbol{B}}$ and the force balance equation becomes $\boldsymbol{n}^{\prime} + \boldsymbol{F}_{L}=\boldsymbol{0}$. We shall write $I\bar{\boldsymbol{B}}=\boldsymbol{B}$ so that the equilibrium equations take the symmetric form
\begin{eqnarray}
\boldsymbol{m}^{\prime} + \boldsymbol{r}^{\prime} \times \boldsymbol{n}=\boldsymbol{0},  \quad\quad  \boldsymbol{n}^{\prime} + \boldsymbol{r}^{\prime} \times \boldsymbol{B}=\boldsymbol{0}, \quad\quad \boldsymbol{B}^{\prime} = \boldsymbol{0}.
\label{eq:magnetic}
\end{eqnarray}
In writing down these equations we have assumed that the current in the rod is moderate so that the effect of the magnetic field generated by the current is negligible compared to the external magnetic field. Eqs \eref{eq:magnetic} have been considered by Wolfe in bifurcation studies from the trivial straight solution \cite{Wolfe88,Seidman88,Wolfe96}. 
%Note that $\boldsymbol{F}_L=\boldsymbol{0}$ if the rod is straight and aligned with the magnetic field. Any perturbation from the straight configuration, however, will result in a force $\boldsymbol{F}_L$ that tends to rotate the rod about the magnetic field.
%
\par
In the director frame the governing equation is a non-canonical Hamiltonian system of the form
\begin{eqnarray}
\left(
\begin{array}{c}
{\mathsf{m}} \\
{\mathsf{n}} \\
{\mathsf{B}} 
\end{array}
\right)^{\prime} & =
{J} \left({\mathsf{m}},{\mathsf{n}},{\mathsf{B}}\right) \nabla \mathcal{H}\left({\mathsf{m}},{\mathsf{n}} \right),
\label{eq:noncanonical_structure}
\end{eqnarray}
where the structure matrix ${J}={J} \left( {\mathsf{m}}, {\mathsf{n}}, {\mathsf{B}}  \right)$ is given by
\begin{eqnarray}
{J} = -{J}^{T} & = \left( 
\begin{array}{ccc}
\hat{{\mathsf{m}}} & \hat{{\mathsf{n}}} & \hat{{\mathsf{B}}} \\
\hat{{\mathsf{n}}} & \hat{{\mathsf{B}}} & {\mathsf{0}} \\
\hat{{\mathsf{B}}} & {\mathsf{0}} & {\mathsf{0}}  
\end{array}
\right)
\label{eq:magnetic_structure}
\end{eqnarray}
and the Hamiltonian is 
\begin{eqnarray}
\mathcal{H} & = \frac{1}{2} \mathsf{m} \cdot \mathsf{u} + \mathsf{d}_{3} \cdot \mathsf{n} . \label{eq:magnetic_ham}
\end{eqnarray}
We note that the Hamiltonian is the same as for the Kirchhoff rod in \eref{eq:kirchhoff_hamiltonian}: the effect of the magnetic field is only present in the structure matrix. The governing equations can be written as
\begin{eqnarray}
\mathsf{m}^{\prime} & = \left\{\mathsf{m}, \mathcal{H} \right\}_{\left({\mathsf{m}},{\mathsf{n}}, {\mathsf{B}} \right)} = \mathsf{m}\times\mathsf{u}+\mathsf{n}\times\mathsf{d}_{3}, \label{eq:magnetic_moment} \\
\mathsf{n}^{\prime} & = \left\{\mathsf{n}, \mathcal{H} \right\}_{\left({\mathsf{m}},{\mathsf{n}}, {\mathsf{B}} \right)} = \mathsf{n}\times\mathsf{u}+\mathsf{B}\times\mathsf{d}_{3}, \label{eq:magnetic_force} \\
\mathsf{B}^{\prime} & = \left\{\mathsf{B}, \mathcal{H} \right\}_{\left({\mathsf{m}},{\mathsf{n}}, {\mathsf{B}} \right)} = \mathsf{B} \times \mathsf{u}, \label{eq:magnetic_vector}
\end{eqnarray}
where the Lie-Poisson bracket on $\left( \mathsf{m}, \mathsf{n}, \mathsf{B} \right)$,
\begin{eqnarray}
\fl
\eqalign{
\left\{ f,g \right\}_{\left({\mathsf{m}},{\mathsf{n}}, {\mathsf{B}} \right)} & =
- {\mathsf{m}}\cdot\left(\nabla_{{\mathsf{m}}}f\times\nabla_{{\mathsf{m}}}g \right) 
- {\mathsf{n}}\cdot\left(\nabla_{{\mathsf{m}}}f\times\nabla_{{\mathsf{n}}}g + \nabla_{{\mathsf{n}}}f\times\nabla_{{\mathsf{m}}}g\right) \nonumber \\
&  \hspace{1.25cm} - \underbrace{{\mathsf{B}} \cdot \left(\nabla_{{\mathsf{m}}}f \times \nabla_{{\mathsf{B}}}g +
\nabla_{{\mathsf{B}}}f \times \nabla_{{\mathsf{m}}}g \right)}_{\mbox{evolution of field}}
- \underbrace{{\mathsf{B}} \cdot \left( \nabla_{{\mathsf{n}}} f \times \nabla_{{\mathsf{n}}} g \right)}_{\mbox{effect of field}},}
\label{eq:magnetic_bracket}
\end{eqnarray}
has been extended from the previous bracket \eref{eq:kirchhoff_bracket} by the addition of two more terms. The first term, due to the evolution of the magnetic field in the director frame, does not affect the governing equations since the Hamiltonian is independent of $\mathsf{B}$. The second term contains all the magnetic effects. This term makes the bracket extension non-semidirect \cite{Thiffeault00}. Note that \eref{eq:magnetic_vector} is merely the trivial equation $\boldsymbol{B}'=\boldsymbol{0}$ rewritten in body components. Eqs \eref{eq:magnetic_moment}-\eref{eq:magnetic_vector} give a physical realisation of the `twisted top' introduced as an abstract mathematical system in \cite{Thiffeault01}.
\par
There are three Casimirs, given by
\begin{eqnarray}
C_{1} & = \frac{1}{2}\mathsf{n} \cdot \mathsf{n} + \mathsf{m} \cdot \mathsf{B}, \label{eq:magnetic_casimirs1} \\
C_{2} & = \mathsf{B} \cdot \mathsf{n}, \label{eq:magnetic_casimirs2} \\
C_{3} & = \mathsf{B} \cdot \mathsf{B}. \label{eq:magnetic_casimirs3}
\end{eqnarray} 
We observe that the magnitude of force is no longer conserved, but as a result of rotational symmetry, the force component in the direction of the magnetic field is conserved resulting in \eref{eq:magnetic_casimirs2}. The magnitude of the magnetic interaction is conserved, thus \eref{eq:magnetic_casimirs3}. Casimir \eref{eq:magnetic_casimirs1} however does not seem to have a physical interpretation.
\par
In the isotropic case ($K_1=K_2$) we have the integrals
%
% \begin{subequations}
\begin{eqnarray}
I_{1} & = K\mathsf{m}\cdot\mathsf{d}_3 \quad\quad (\mbox{if}~~K_{1} = K_{2} = K), \label{eq:magnetic_lagrange} \\
I_{2} & = {\mathsf{n}} \cdot {\mathsf{m}} + K {\mathsf{B}} \cdot {\mathsf{d}}_{3} \quad\quad (\mbox{if}~~K_{1} = K_{2}=K). \label{eq:int2}
\end{eqnarray}
% \end{subequations}
%
The first of these expresses conservation of twist in the rod, as in the Kirchhoff case. The second integral, like does not seem to have a physical interpretation when $C_{3}\ne0$. We note that the Lagrange integrability condition $K_1=K_2$ is unaltered by the magnetic field. The authors in \cite{Thiffeault01} give numerical evidence (in the form of chaotic orbits) that the same is not true for the Kovalevskaya case: the magnetic rod with $K_1=K_3=2K_2$ is not integrable. Of course, a $\boldsymbol{B}$-perturbed condition on the stiffnesses may exist for which the system is integrable.
\par
It is a straightforward task to check that all the integrals \eref{eq:magnetic_ham}, \eref{eq:magnetic_lagrange} and \eref{eq:int2} are independent and in involution with respect to the Lie-Poisson bracket \eref{eq:magnetic_bracket}. Hence in the isotropic case the system is completely integrable in the sense of Arnold-Liouville. In the next section we use the conserved quantities to reduce the nine-dimensional system \eref{eq:magnetic_moment}-\eref{eq:magnetic_vector}. 
% In principle action-angle coordinates exist which reduce the Hamiltonian to a five-tori defined by the Hamiltonian $\mathcal{H}=h$, the two first integrals and two of the Casimirs. 
%
\section{Reduction of the magnetic rod} \label{sec:reduction} 
In this section we use the three Casimirs \eref{eq:magnetic_casimirs1}-\eref{eq:magnetic_casimirs3} to reduce the nine-dimensional non-canonical Hamiltonian system to a six-dimensional canonical Hamiltonian system. This is possible (at least locally) provided the structure matrix $J$ is of constant rank everywhere \cite[Section 6.2]{Olver93}. We perform the reduction by constructing a coordinate transformation from the nine coordinates $(\mathsf{m},\mathsf{n},\mathsf{B}$ to three Euler angles $q=(\theta,\psi,\phi)$ and their canonical momenta $p=(p_\theta,p_\psi,p_\phi)$. In the isotropic case the system reduces further to a four-dimensional integrable canonical Hamiltonian system. We follow \cite{Thiffeault01} but give details here to show the system is canonical. As it happens, the transformation is only canonical subject to a non-alignment condition. The aligned case is also of interest and is treated in subsection \ref{sec:alignment}. A reduction similar to the one in the present section was carried out in \cite{Mielke88} for the nonintegrable anisotropic free rod.
\par
Let
\begin{eqnarray}
\fl
R & = 
\left(
\begin{array}{ccc}
\cos\theta\cos\phi\cos\psi-\sin\phi\sin\psi & \cos\theta\cos\phi\sin\psi+\cos\psi\sin\phi & -\sin\theta\cos\phi \\
-\cos\theta\sin\phi\cos\psi-\cos\phi\sin\psi & -\cos\theta\sin\phi\sin\psi+\cos\phi\cos\psi & \sin\theta\sin\phi \\
\sin\theta\cos\psi & \sin\theta\sin\psi & \cos\theta
\end{array}
\right)
\nonumber
\end{eqnarray}
be a parametrisation of the rotation matrix $R$ in \eref{eq:frame} in terms of Euler angles. Here we have followed the convention used, e.g., by Love \cite{Love44}, thus $\theta$ is the angle the tangent to the rod makes with the magnetic field, $\psi$ is the azimuthal angle about the field direction and $\phi$ is the twist angle about the centreline of the rod. We may assume, without loss of generality, that $\boldsymbol{B}$ is directed along the $\boldsymbol{e}_3$ vector of the fixed coordinate system. For the triple $\mathsf{B}$ we then have
\begin{eqnarray}
\mathsf{B}\left(q\right) & = \sqrt{C_{3}} R\left(q\right) k = \sqrt{C_{3}}
\left(
\begin{array}{c}
-\sin\theta\cos\phi\\
\sin\theta\sin\phi\\
\cos\theta
\end{array}
\right),
\label{eq:canonical_magnetic}
\end{eqnarray}
where $k=\left(0,0,1\right)^T$. On inserting the Euler angles into the strains \eref{eq:curvatures} and using the constitutive relations \eref{eq:constit} we obtain
\begin{eqnarray}
\mathsf{m} & =
\left(
\begin{array}{c}
m_1 \\
m_2 \\
m_3
\end{array}
\right)
=
\left( 
\begin{array}{c}
K_1(\theta^{\prime} \sin\phi - \psi^{\prime} \sin\theta \cos\phi) \\
K_2(\theta^{\prime} \cos\phi + \psi^{\prime} \sin\theta \sin\phi) \\
K_3(\phi^{\prime} + \psi^{\prime} \cos\theta)
\end{array}
\right)
=Lp,
\label{eq:canonical_moment}
\end{eqnarray}
where
\begin{eqnarray}
L & = \frac{1}{\sin\theta} 
\left( 
\begin{array}{ccc}
\sin\theta\sin\phi & -\cos\phi &  \cos\theta\cos\phi \\
\sin\theta\cos\phi &  \sin\phi & -\cos\theta\sin\phi \\
0 & 0 & \sin\theta 
\end{array}
\right) \quad \mbox{and} \quad p=(p_\theta,p_\psi,p_\phi) 
\nonumber
\end{eqnarray}
are the canonical momenta defined by $p_\theta=\partial \overline W(q,q')/\partial \theta'$,
$p_\psi=\partial \overline W(q,q')/\partial \psi'$, $p_\phi=\partial \overline W(q,q')/\partial \phi'$, with $\overline W(q,q')=W(\mathsf{u}(q,q'))$ in terms of the strain energy function $W$ defined in \eref{eq:strain_energy}. Finally, for the force we may write
\begin{eqnarray}
\mathsf{n}\left( q,p \right) & = R\left(q\right)v\left(q,p\right), \nonumber
\end{eqnarray}
for some non-constant triple $v$. Decomposing $v$ as
\begin{eqnarray}
v\left(q,p\right) & = v_{\perp}\left(q,p\right) {i}_{\perp} + v_{\parallel}\left(q,p\right) {i}_{\parallel}, \nonumber
\end{eqnarray}
where ${i}_{\parallel}$ and ${i}_{\perp}$ are unit triples parallel and perpendicular to $k$, respectively, we obtain
\begin{eqnarray}
C_{2} & = \mathsf{B} \cdot \mathsf{n}=\sqrt{C_{3}} R k \cdot R v = \sqrt{C_{3}} v \cdot \left( R^{T} R \right) k = \sqrt{C_{3}} v_{\parallel}.
\end{eqnarray}
Furthermore,
\begin{eqnarray}
C_{1} & = \frac{1}{2}R v\cdot Rv + \sqrt{C_{3}} L p \cdot R k = \frac{C_{2}^{2}}{2C_{3}} + \frac{1}{2}v_{\perp}^{2} + \sqrt{C_{3}} p \cdot \left( L R^{T} \right) k,
\end{eqnarray}
which allows us to solve for $v_{\perp}$:
\begin{eqnarray}
\fl
v_{\perp}\left( q,p \right) & = \sqrt{ 2C_{1}^{} - \frac{C_{2}^{2}}{C_{3}} - 2 \sqrt{C_{3}} p \cdot \left( L R^{T} \right) k} =
\sqrt{ 2C_{1}^{} - \frac{C_{2}^{2}}{C_{3}} - 2 \sqrt{C_{3}}\,p_{\psi}^{} },
\label{eq:handedness}
\end{eqnarray}
where, without loss of generality, we have taken the positive solution. If the vector perpendicular to $k$ is taken to be $i_{\perp} = \left( 1,0,0 \right)^T$ we obtain
\begin{eqnarray}
% \fl
\mathsf{n} & = 
\frac{C_{2}^{}}{\sqrt{C_{3}}}
\left( 
\begin{array}{c}
-\sin\theta\cos\phi \\
\sin\theta\sin\phi \\
\cos\theta
\end{array}
\right) \nonumber \\
& \hspace{1.0cm}
+ \sqrt{ 2C_{1}^{} - \frac{C_{2}^{2}}{C_{3}} - 2 \sqrt{C_{3}}\,p_{\psi}^{}  }
\left(\begin{array}{c}
 \cos\theta\cos\phi\cos\psi-\sin\phi\sin\psi \\
-\cos\theta\sin\phi\cos\psi-\cos\phi\sin\psi \\
 \sin\theta\cos\psi
\end{array}
\right). \label{eq:canonical_force}
\end{eqnarray}
Note that this transformation is well defined as $2C_{1}^{} - \frac{C_{2}^{2}}{C_{3}} - 2 \sqrt{C_{3}}p_{\psi}^{}=v_{\perp}^{2} \ge 0$.
\par
Eqs \eref{eq:canonical_magnetic}, \eref{eq:canonical_moment} and \eref{eq:canonical_force} give the sought transformation. The Jacobian matrix of its inverse is
\begin{eqnarray}
G & = \frac{ \partial \left( q, p \right) }{ \partial \left( \mathsf{m}, \mathsf{n}, \mathsf{B} \right) }\cdot
\label{eq:G}
\end{eqnarray}
In order for the bracket \eref{eq:magnetic_bracket} to be transformed to canonical form we need to verify that
\begin{eqnarray}
G J G^{T} & = \bar{J}, \nonumber
\end{eqnarray}
where $J$ is the structure matrix defined in \eref{eq:magnetic_structure} and $\bar{J}$ is the standard canonical structure matrix in $\mathbb{R}^{6}$. We show in the Appendix that this is indeed the case, {\em provided} that $v_{\perp}>0$, i.e., provided that $\mathsf{n}$ and $\mathsf{B}$ are not aligned. Without this condition the necessary inverse transformation does not exist. Note that $\mathsf{n}$ and $\mathsf{B}$ are aligned if and only if $2C_1^{}-\frac{C_2^2}{C_{3}}=2 \mathsf{m}\cdot\mathsf{B}$. Now from \eref{eq:magnetic_casimirs1}, \eref{eq:magnetic_force} and the conservation of $C_1$ we have
\begin{eqnarray}
2\frac{\mbox d}{\mbox{d}s}(\mathsf{m}\cdot\mathsf{B})=-\frac{\mbox d}{\mbox{d}s}(\mathsf{n}\cdot\mathsf{n})=2\mathsf{d}_3\cdot(\mathsf{B}\times\mathsf{n}),
\label{eq:align}
\end{eqnarray}
which vanishes if $\mathsf{n}$ and $\mathsf{B}$ are aligned. Thus the alignment condition is well defined: if the force and the magnetic field are aligned anywhere they are aligned everywhere along the rod.
\par
The Hamiltonian \eref{eq:magnetic_ham} is transformed to
\begin{eqnarray}
\fl
\lefteqn{\mathcal{H} = \frac{1}{2K_1K_2\sin^2\theta}\left[p_\theta^2\,\sin^2\theta\,(K_2-
(K_2-K_1)\cos^2\phi)+p_\psi^2\,(K_1\sin^2\phi+K_2\cos^2\phi) \right.}
\nonumber \\
&& + p_\phi^2\left(\cos^2\theta\,(K_1\sin^2\phi+K_2\cos^2\phi)
+(K_1K_2/K_3)\sin^2\theta\right) \nonumber\\
&& + 2(K_2-K_1)p_\theta p_\phi\sin\theta\cos\theta\sin\phi\cos\phi -
2(K_2-K_1)p_\theta p_\psi\sin\theta\sin\phi\cos\phi \\
&& \left.- 2p_\psi p_\phi\cos\theta\left(K_1\sin^2\phi+K_2\cos^2\phi\right)
\right] \nonumber \\
&& + \frac{C_{2}^{}}{\sqrt{C_{3}}} \cos\theta + \sin\theta\cos\psi \sqrt{ 2C_{1}^{} - \frac{C_{2}^{2}}{C_{3}} - 2 \sqrt{C_{3}}\,p_{\psi}^{} }. \label{eq:ham_canon} 
\end{eqnarray}
Note that the effect of the magnetic field, encoded by $C_3$, has been transferred from the Lie-Poisson bracket to the Hamiltonian. The limit $C_3\to 0$ is singular, as was also observed in \cite{Thiffeault01}.
\subsection{The isotropic case}
In the isotropic case ($K_1=K_2=K$) the Hamiltonian \eref{eq:ham_canon} reduces to
\begin{eqnarray}
% \fl
\mathcal{H} & = \frac{p_{\theta}^{2}}{2 K} + \frac{ \left( p_{\psi}^{}-p_{\phi}^{}\cos\theta \right)^{2} }{2 K \sin^{2}\theta} + \frac{1}{2 K_3}p_{\phi}^{2} +  \frac{C_{2}^{}}{\sqrt{C_{3}}} \cos\theta  \nonumber \\
& \hspace{1.50cm} + \sin\theta\cos\psi \sqrt{ 2C_{1}^{} - \frac{C_{2}^{2}}{C_{3}} - 2 \sqrt{C_{3}}\,p_{\psi}^{} }. \label{eq:Euler_Ham}
\end{eqnarray}
Since this Hamiltonian does not depend on the angle $\phi$ the momentum $p_\phi=m_3$ is a constant. We also have the additional integral \eref{eq:int2}, which in canonical variables reads
\begin{eqnarray}
\mathcal{I} & = \sqrt{C_{3}} K \cos\theta + \frac{C_{2}^{}}{\sqrt{C_{3}}} p_{\psi}  \nonumber \\
& \hspace{0.5cm} - \sqrt{ 2C_{1}^{} - \frac{C_{2}^{2}}{C_{3}}2 - \sqrt{C_{3}}\,p_{\psi}^{} } \left( p_{\theta}\sin\psi - \cos\psi \left( \frac{p_{\phi} - p_{\psi}\cos\theta}{\sin\theta} \right) \right),
\label{eq:constraint}
\end{eqnarray}
rendering the system completely integrable.
\par
Hamilton's equations are
%
% \begin{subequations}
\label{eq:all}
\begin{eqnarray}
% \fl
{\theta}^{\prime}  & = \frac{p_{\theta}}{K}, \label{eq:ham3}\\ 
% \fl
{\psi}^{\prime}  & =  \frac{ \left( p_{\psi} - p_{\phi} \cos\theta \right) }{ K \sin^{2}\theta }
- \frac{ \sqrt{C_{3}} \cos\psi\sin\theta }{ \sqrt{2C_{1}^{} - \frac{C_{2}^{2}}{C_{3}} - 2 \sqrt{C_{3}}\,p_{\psi}^{} } }, \label{eq:ham4}\\
% \fl
{p}_{\theta}^{\prime} & = \frac{\left( p_{\psi}^{}\cos\theta - p_{\phi}^{} \right)\left( p_{\psi}^{} - p_{\phi}^{}\cos\theta \right)}{ K \sin^{3}\theta } + \frac{C_{2}}{\sqrt{C_{3}}} \sin\theta \nonumber \\
% \fl
& \hspace{2.5cm} - \cos\theta\cos\psi\sqrt{ 2C_{1}^{} - \frac{C_{2}^{2}}{C_{3}} - 2 \sqrt{C_{3}}\,p_{\psi}^{} },  \label{eq:ham1} \\
% \fl
{p}_{\psi}^{\prime}  & = \sin\theta\sin\psi\sqrt{ 2C_{1}^{} - \frac{C_{2}^{2}}{C_{3}} - 2 \sqrt{C_{3}}p_{\psi}^{} } \label{eq:ham2}.
\end{eqnarray}
% \end{subequations}
%
Helical solutions about $\boldsymbol{B}=\boldsymbol{e}_3$ have $\theta=\mbox{const.}$, $\psi^{\prime}=\mbox{const.}$ By integrating \eref{eq:ham2} subject to these conditions and inserting the result into \eref{eq:ham4} one can show that no helical solutions about $\boldsymbol{e}_3$ exist in this non-aligned case.
\par
\begin{figure}[h!tb]
\label{fig:phase}
\begin{center}
\subfigure[][$\cos\psi=0.9$]{ \begin{picture}(0,0)%
\includegraphics{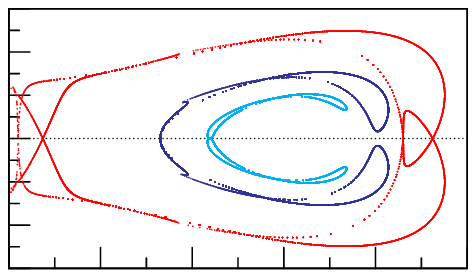}%
\end{picture}%
\begingroup
\setlength{\unitlength}{0.0200bp}%
\begin{picture}(9900,5940)(0,0)%
\put(2200,1650){\makebox(0,0)[r]{\strut{}-1.5}}%
\put(2200,2273){\makebox(0,0)[r]{\strut{}-1}}%
\put(2200,2897){\makebox(0,0)[r]{\strut{}-0.5}}%
\put(2200,3520){\makebox(0,0)[r]{\strut{} 0}}%
\put(2200,4143){\makebox(0,0)[r]{\strut{} 0.5}}%
\put(2200,4767){\makebox(0,0)[r]{\strut{} 1}}%
\put(2200,5390){\makebox(0,0)[r]{\strut{} 1.5}}%
\put(2475,1100){\makebox(0,0){\strut{} 0}}%
\put(3795,1100){\makebox(0,0){\strut{} 0.5}}%
\put(5115,1100){\makebox(0,0){\strut{} 1}}%
\put(6435,1100){\makebox(0,0){\strut{} 1.5}}%
\put(7755,1100){\makebox(0,0){\strut{} 2}}%
\put(9075,1100){\makebox(0,0){\strut{} 2.5}}%
\put(550,3520){\rotatebox{90}{\makebox(0,0){\strut{}$p_{\theta}^{}$}}}%
\put(5775,275){\makebox(0,0){\strut{}$\theta$}}%
\end{picture}%
\endgroup 
\label{fig:1a} }
\subfigure[][$\cos\psi=0.7$]{  \begin{picture}(0,0)%
\includegraphics{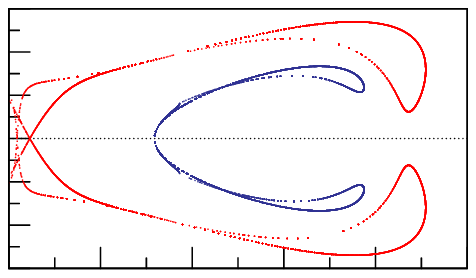}%
\end{picture}%
\begingroup
\setlength{\unitlength}{0.0200bp}%
\begin{picture}(9900,5940)(0,0)%
\put(2200,1650){\makebox(0,0)[r]{\strut{}-1.5}}%
\put(2200,2273){\makebox(0,0)[r]{\strut{}-1}}%
\put(2200,2897){\makebox(0,0)[r]{\strut{}-0.5}}%
\put(2200,3520){\makebox(0,0)[r]{\strut{} 0}}%
\put(2200,4143){\makebox(0,0)[r]{\strut{} 0.5}}%
\put(2200,4767){\makebox(0,0)[r]{\strut{} 1}}%
\put(2200,5390){\makebox(0,0)[r]{\strut{} 1.5}}%
\put(2475,1100){\makebox(0,0){\strut{} 0}}%
\put(3795,1100){\makebox(0,0){\strut{} 0.5}}%
\put(5115,1100){\makebox(0,0){\strut{} 1}}%
\put(6435,1100){\makebox(0,0){\strut{} 1.5}}%
\put(7755,1100){\makebox(0,0){\strut{} 2}}%
\put(9075,1100){\makebox(0,0){\strut{} 2.5}}%
\put(550,3520){\rotatebox{90}{\makebox(0,0){\strut{}$p_{\theta}^{}$}}}%
\put(5775,275){\makebox(0,0){\strut{}$\theta$}}%
\end{picture}%
\endgroup 
\label{fig:1b} }
\subfigure[][$\cos\psi=0.5$]{ \begin{picture}(0,0)%
\includegraphics{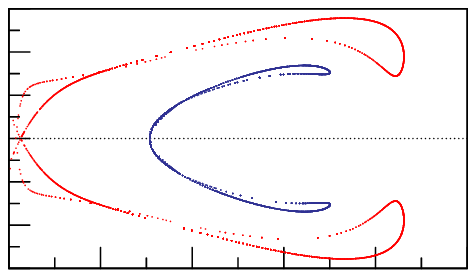}%
\end{picture}%
\begingroup
\setlength{\unitlength}{0.0200bp}%
\begin{picture}(9900,5940)(0,0)%
\put(2200,1650){\makebox(0,0)[r]{\strut{}-1.5}}%
\put(2200,2273){\makebox(0,0)[r]{\strut{}-1}}%
\put(2200,2897){\makebox(0,0)[r]{\strut{}-0.5}}%
\put(2200,3520){\makebox(0,0)[r]{\strut{} 0}}%
\put(2200,4143){\makebox(0,0)[r]{\strut{} 0.5}}%
\put(2200,4767){\makebox(0,0)[r]{\strut{} 1}}%
\put(2200,5390){\makebox(0,0)[r]{\strut{} 1.5}}%
\put(2475,1100){\makebox(0,0){\strut{} 0}}%
\put(3795,1100){\makebox(0,0){\strut{} 0.5}}%
\put(5115,1100){\makebox(0,0){\strut{} 1}}%
\put(6435,1100){\makebox(0,0){\strut{} 1.5}}%
\put(7755,1100){\makebox(0,0){\strut{} 2}}%
\put(9075,1100){\makebox(0,0){\strut{} 2.5}}%
\put(550,3520){\rotatebox{90}{\makebox(0,0){\strut{}$p_{\theta}^{}$}}}%
\put(5775,275){\makebox(0,0){\strut{}$\theta$}}%
\end{picture}%
\endgroup
\label{fig:1c} }
\subfigure[][$\cos\psi=0.3$]{ \begin{picture}(0,0)%
\includegraphics{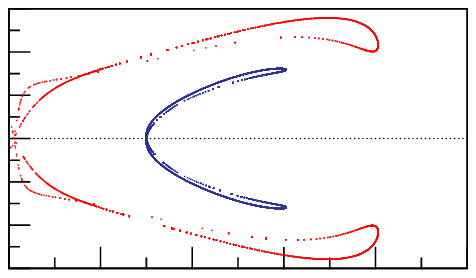}%
\end{picture}%
\begingroup
\setlength{\unitlength}{0.0200bp}%
\begin{picture}(9900,5940)(0,0)%
\put(2200,1650){\makebox(0,0)[r]{\strut{}-1.5}}%
\put(2200,2273){\makebox(0,0)[r]{\strut{}-1}}%
\put(2200,2897){\makebox(0,0)[r]{\strut{}-0.5}}%
\put(2200,3520){\makebox(0,0)[r]{\strut{} 0}}%
\put(2200,4143){\makebox(0,0)[r]{\strut{} 0.5}}%
\put(2200,4767){\makebox(0,0)[r]{\strut{} 1}}%
\put(2200,5390){\makebox(0,0)[r]{\strut{} 1.5}}%
\put(2475,1100){\makebox(0,0){\strut{} 0}}%
\put(3795,1100){\makebox(0,0){\strut{} 0.5}}%
\put(5115,1100){\makebox(0,0){\strut{} 1}}%
\put(6435,1100){\makebox(0,0){\strut{} 1.5}}%
\put(7755,1100){\makebox(0,0){\strut{} 2}}%
\put(9075,1100){\makebox(0,0){\strut{} 2.5}}%
\put(550,3520){\rotatebox{90}{\makebox(0,0){\strut{}$p_{\theta}^{}$}}}%
\put(5775,275){\makebox(0,0){\strut{}$\theta$}}%
\end{picture}%
\endgroup 
\label{fig:1d} }
\end{center}
\caption{Poincar\'e plots for \eref{eq:ham3}-\eref{eq:ham2} with various sections. In each diagram orbits are displayed for energy levels $\mathcal{H}=1.90$, 1.50 and 1.37, while $\mathcal{I}= 1.00995$, $\lambda=0.01$, $C_1=1.02$, $C_2=C_3=p_\phi=1$ and $K_3/K=3/4$.}
\end{figure}

The phase space of the reduced system \eref{eq:ham3}-\eref{eq:ham2} can be explored by means of (planar projections of) two-dimensional Poincar\'e sections of level sets of the integrals $\mathcal{I}$ and $\mathcal{H}$. Fig.~\ref{fig:phase} shows examples in which we have fixed $\mathcal{I}=1.00995$ and recorded intersections of a few orbits with planes of section given by $\cos\psi=\alpha$, for various values of $\alpha$.
\subsection{Alignment of force and field -- The superintegrable case} \label{sec:alignment}
Here we consider the physically realistic case that the force and the magnetic field are aligned. We already showed that if force and field are aligned anywhere then they are aligned everywhere. From \eref{eq:magnetic} it follows that $\boldsymbol{d}_3\times\boldsymbol{B}=\boldsymbol{0}=\boldsymbol{d}_3\times\boldsymbol{n}$, i.e., $\boldsymbol{n}$ is aligned with $\boldsymbol{d}_3$, while also $\boldsymbol{n}$ and $\boldsymbol{m}$ are constant. This means that solutions are twisted straight rods. Hence the aligned case is maximally superintegrable with solutions lying on one-tori. Note that this conclusion holds irrespective of whether the rod is isotropic.
\section{Generalised magnetic rods} \label{sec:generalized}
\subsection{A rod in a nonuniform magnetic field -- Four-field model} \label{subsec:twisted_field}
By inspection of the structure matrices \eref{eq:twist_structure}, \eref{eq:kirchhoff_structure} and \eref{eq:magnetic_structure} we may go to the next member of the hierarchy and consider the equations
\begin{eqnarray}
\boldsymbol{m}^{\prime} + \boldsymbol{r}^{\prime}\times\boldsymbol{n} = \boldsymbol{0}, \quad \boldsymbol{n}^{\prime} + \boldsymbol{r}^{\prime}\times\boldsymbol{B}=\boldsymbol{0}, \quad \boldsymbol{B}^{\prime} + \boldsymbol{r}^{\prime}\times\boldsymbol{D}=\boldsymbol{0}, \quad \boldsymbol{D}^{\prime} = \boldsymbol{0}.
\label{eq:nonuniform_vec}
\end{eqnarray}
The equation for $\boldsymbol{B}$ can be integrated to give $B_x=y$, $B_y=-x$, $B_z=0$, where $x,y,z$ and $B_x,B_y,B_z$ are components of $\boldsymbol{r}$ and $\boldsymbol{B}$ relative to the fixed frame $\{\boldsymbol{e}_1,\boldsymbol{e}_2,\boldsymbol{e}_3\}$, and we have chosen $\boldsymbol{e}_3$ in the direction of $\boldsymbol{D}$. Thus eqs \eref{eq:nonuniform_vec} can be thought of as describing a rod in a linearly-varying magnetic field generated by a uniform `hypermagnetic' field $\boldsymbol{D}$.
\par
In the director frame the equations take the Hamiltonian form
\begin{eqnarray}
\left( \begin{array}{cccc} 
\mathsf{m} \\
\mathsf{n} \\
\mathsf{B} \\
\mathsf{D}
\end{array} \right)^{\prime} & = J\left( \mathsf{m}, \mathsf{n}, \mathsf{B}, \mathsf{D} \right) \nabla \mathcal{H}\left( \mathsf{m},\mathsf{n}\right), \quad \mbox{with} \quad \mathcal{H}\left(\mathsf{m},\mathsf{n}\right) = \frac{1}{2}\mathsf{u}\cdot\mathsf{m} + \mathsf{d}_{3}\cdot\mathsf{n}, \nonumber
\end{eqnarray}
and structure matrix
\begin{eqnarray}
J = -J^{T} & = \left( 
\begin{array}{cccc} 
\hat{\mathsf{m}} & \hat{\mathsf{n}} & \hat{\mathsf{B}} & \hat{\mathsf{D}} \\
\hat{\mathsf{n}} & \hat{\mathsf{B}} & \hat{\mathsf{D}} & \mathsf{0} \\
\hat{\mathsf{B}} & \hat{\mathsf{D}} & \mathsf{0} & \mathsf{0} \\
\hat{\mathsf{D}} & \mathsf{0} & \mathsf{0} & \mathsf{0}
\end{array} 
\right).
\end{eqnarray}
Or we can write
%
% \begin{subequations}
\label{eq:nonuniform}
\begin{eqnarray}
\mathsf{m}^{\prime} & = \left\{ \mathsf{m}, \mathcal{H} \right\}_{\left({\mathsf{m}},{\mathsf{n}},{\mathsf{B}}, {\mathsf{D}} \right)} = \mathsf{m} \times \mathsf{u} + \mathsf{n} \times \mathsf{d}_{3}, \\
\mathsf{n}^{\prime} & = \left\{ \mathsf{n}, \mathcal{H} \right\}_{\left({\mathsf{m}},{\mathsf{n}},{\mathsf{B}}, {\mathsf{D}} \right)} = \mathsf{n} \times \mathsf{u} + \mathsf{B} \times \mathsf{d}_{3}, \\
\mathsf{B}^{\prime} & = \left\{ \mathsf{B}, \mathcal{H} \right\}_{\left({\mathsf{m}},{\mathsf{n}},{\mathsf{B}}, {\mathsf{D}} \right)} = \mathsf{B} \times \mathsf{u} + \mathsf{D} \times \mathsf{d}_{3}, \\
\mathsf{D}^{\prime} & = \left\{ \mathsf{D}, \mathcal{H} \right\}_{\left({\mathsf{m}},{\mathsf{n}},{\mathsf{B}}, {\mathsf{D}} \right)} =\mathsf{D} \times \mathsf{u},
\end{eqnarray}
% \end{subequations}
%
where the Lie-Poisson bracket is constructed from \eref{eq:magnetic_bracket} through the addition of another semidirect and non-semidirect extension:
\begin{eqnarray}
\fl
\left\{ f,g \right\}_{\left({\mathsf{m}},{\mathsf{n}},{\mathsf{B}}, {\mathsf{D}} \right)} & =
- {\mathsf{m}}\cdot\left(\nabla_{{\mathsf{m}}}f\times\nabla_{{\mathsf{m}}}g \right) 
- {\mathsf{n}}\cdot\left(\nabla_{{\mathsf{m}}}f\times\nabla_{{\mathsf{n}}}g + \nabla_{{\mathsf{n}}}f\times\nabla_{{\mathsf{m}}}g\right) \nonumber \\
\fl
&  \hspace{2.0cm} {}- {\mathsf{B}} \cdot \left(\nabla_{{\mathsf{m}}}f \times \nabla_{{\mathsf{B}}}g +
\nabla_{{\mathsf{B}}}f \times \nabla_{{\mathsf{m}}}g \right)
- {\mathsf{B}} \cdot \left( \nabla_{{\mathsf{n}}} f \times \nabla_{{\mathsf{n}}} g \right) \nonumber \\
\fl
&  \hspace{2.0cm} {}-\underbrace{ {\mathsf{D}} \cdot \left(\nabla_{{\mathsf{m}}}f \times \nabla_{{\mathsf{D}}}g +
\nabla_{{\mathsf{D}}}f \times \nabla_{{\mathsf{m}}}g \right)}_{\mbox{evolution of hyperfield}}
- \underbrace{{\mathsf{D}} \cdot \left( \nabla_{{\mathsf{B}}} f \times \nabla_{{\mathsf{n}}} g \right)}_{\mbox{effect of hyperfield}}. \nonumber
\end{eqnarray}
\par
This twelve-dimensional system has four independent Casimirs:
%
% \begin{subequations}
% \label{eq:nonuniform_casimirs}
\begin{eqnarray}
C_{1} & = \mathsf{m}\cdot\mathsf{D} + \mathsf{n}\cdot\mathsf{B}, \label{eq:nonuniform_casimir_1} \\
C_{2} & = {\frac{1}{2}}\mathsf{B}\cdot\mathsf{B} + \mathsf{n}\cdot\mathsf{D}, \label{eq:nonuniform_casimir_2} \\
C_{3} & = \mathsf{B}\cdot\mathsf{D}, \label{eq:nonuniform_casimir_3}\\
C_{4} & = \mathsf{D}\cdot\mathsf{D}.\label{eq:nonuniform_casimir_4}
\end{eqnarray}
% \end{subequations}
%
In the isotropic case there are now three independent first integrals besides the Hamiltonian,
%
% \begin{subequations}
\label{eq:nonuniform_first_integrals}
\begin{eqnarray}
I_{1} & = {K} \mathsf{m}\cdot\mathsf{d}_{3} \quad\quad (\mbox{if}~~K_{1} = K_{2} = K), \label{eq:nonuniform_integral_1} \\
I_{2} & = \mathsf{n}\cdot\mathsf{m} + K\mathsf{B}\cdot\mathsf{d}_{3} \quad\quad (\mbox{if}~~K_{1} = K_{2}=K), \label{eq:nonuniform_integral_2}\\
I_{3} & = {\frac{1}{2}}\mathsf{n}\cdot\mathsf{n} + \mathsf{m}\cdot\mathsf{B} + K\mathsf{D}\cdot\mathsf{d}_{3} \quad\quad (\mbox{if}~~K_{1} = K_{2}=K),\label{eq:nonuniform_integral_3}
\end{eqnarray}
% \end{subequations}
%
making the system completely integrable. If $C_4 = 0$ then $\mathsf{D}=\mathsf{0}$ and the system reduces to that of the magnetic rod in the previous section. The system loses rank as the Casimir $C_{4}=0$ necessarily implies $C_{3}=0$ and the two Casimirs lose their independent meaning. Interestingly, the integral $I_3$ then becomes a Casimir (cf.~\eref{eq:magnetic_casimirs1}), whose preservation does not rely on isotropy anymore. By using the four Casimirs \eref{eq:nonuniform_casimir_1}-\eref{eq:nonuniform_casimir_4} the twelve-dimensional system can in principle be reduced to an eight-dimensional canonical system.

Alignment again defines a special case. It can be shown in the same way as in the previous section that if $\boldsymbol{D}$ and $\boldsymbol{B}$
are aligned anywhere then they are aligned everywhere. From \eref{eq:nonuniform_vec} it then follows that $\boldsymbol{D}$ and $\boldsymbol{B}$ are aligned with $\boldsymbol{d}_3$, which is therefore constant. Thus all solutions are twisted straight rods. Again we find that alignment leads to a maximally superintegrable case with solutions existing on one-tori.

\subsection{A Lax pair for the isotropic case} \label{subsec:lax_pair}
The previous sections show a hierarchy of rod models based on the form of the structure matrices. Conditions on the constitutive relations determine if the model is integrable. Here we give a compact Lax pair formulation of the integrable hierarchy of isotropic rod problems. Consider the parametrised Lax pair
\begin{eqnarray}
\frac{\mathrm{d}}{\mathrm{d}s} \Gamma\left( \mu \right) & = \left[ \Gamma\left(\mu\right), \hat{\mathsf{d}}_{3}  \mu + \hat{\mathsf{u}} \right], \label{eq:lax_pair}
\end{eqnarray}
where 
\begin{eqnarray}
\Gamma\left(\mu\right) & = K\hat{\mathsf{d}}_{3} \mu + \Gamma_{0} + \Gamma_{1}\mu^{-1} +{} \ldots {}+ \Gamma_{n} \mu^{-n} \in \mathfrak{so}\left(3\right), \quad n \in \mathbb{N},\nonumber
\end{eqnarray}
with 
\begin{eqnarray}
\hat{\mathsf{d}}_{3}  & = 
\left( 
\begin{array}{ccc} 
0 & -1 & 0 \\
1 & 0 & 0 \\
0 & 0 & 0
\end{array} 
\right)
\quad \mbox{and} \quad
\hat{\mathsf{u}} = 
\left( 
\begin{array}{ccc} 
0 & -u_{3} & u_{2} \\
u_{3} & 0 & -u_{1} \\
-u_{2} & u_{1} & 0
\end{array}
\right). \nonumber
\end{eqnarray}
\par
This Lax pair was proposed in \cite{vivolo03} to study monodromy of a generalised family of symmetric Lagrange tops. It describes our hierarchy of rod models if we associate the terms in the expansion of $\Gamma$ by $\mu$ with our field variables: $\Gamma_{0} = {\hat{\mathsf{m}}}$, $\Gamma_{1} = {\hat{\mathsf{n}}}$, $\Gamma_{2} = \hat{\mathsf{B}}$, $\Gamma_{3} = \hat{\mathsf{D}}$, etc. The non-canonical equations for the force-free rod ($n=0$), Kirchhoff rod ($n=1$), rod in uniform magnetic field ($n=2$) and rod in nonuniform magnetic field ($n=3$) are obtained by equating like powers of $\mu$ in \eref{eq:lax_pair}, while the first integrals are generated by 
\begin{eqnarray}
I_{i} & = -\frac{1}{4}\mathrm{residue}_{\mu=0}\left( \mu^{i-1}\mathrm{trace}\left[ \Gamma\left(\mu\right)^{2} \right] \right), \quad \mbox{for} \quad i=-1,0,1,\ldots, n-1, \nonumber
\end{eqnarray}
the Casimirs by
\begin{eqnarray}
C_{i} & = -\frac{1}{4}\mathrm{residue}_{\mu=0}\left( \mu^{i-1}\mathrm{trace}\left[ \Gamma\left(\mu\right)^{2} \right] \right), \quad \mbox{for} \quad i=n,n+1,n+2,\ldots, 2n, \nonumber
\end{eqnarray}
and the Hamiltonian is given by
\begin{eqnarray}
\mathcal{H} & = \frac{I_0}{K} + \frac{ K - K_3 }{2 KK_3}\left(\frac{I_{-1}}{K}\right)^{2}. \nonumber
\end{eqnarray}
\section{Conclusion} \label{sec:conc}
We have shown the equations for a conducting rod in a uniform magnetic field to sit (as third member) in a hierarchy of rod models, the first two of which are the classical force-free rod and the Kirchhoff rod. The hierarchy can be interpreted as describing a succession of generalised body forces that are all of the form $\boldsymbol{F}\times\boldsymbol{d}_{3}$ for some field $\boldsymbol{F}$, i.e., they act normal to the rod. In the case of a transversely isotropic unshearable and inextensible rod the equilibrium equations are found to be completely integrable. The integrable hierarchy of isotropic rod models is generated by a parametrised Lax pair. The integrable models also have maximally superintegrable cases which allow solutions to be derived by purely algebraic means, giving straight twisted configurations for the rod.
\par
The uniform magnetic rod gives a physical realisation to the abstract `twisted top' equations introduced in \cite{Thiffeault01}. The fourth member of the hierarchy also has a physical interpretation as a conducting rod in a non-uniform linearly-varying magnetic field. The next member is not so easy to interpret as the magnetic field $\boldsymbol{B}$ depends on the configuration of the rod; it is no longer an external field. Not all the conserved quantities have a physical interpretation, for example \eref{eq:magnetic_casimirs1} and \eref{eq:int2} in the three-field model. This is not unusual: the Kovalevskaya integral \eref{eq:kov} for the heavy top (or Kirchhoff rod), for instance, does not have a physical interpretation. 
\par
Unlike the Lagrange case the Kovalevskaya integrable case is destroyed by the perturbation due to the magnetic field, although a Kovalevskaya type case may exist at a different condition. A hierarchy of generalised Kovalevskaya tops is given in \cite{Bobenko89}, where also a parametrised Lax pair for this top is derived. Here the perturbation is a body moment rather than a body force. The Lax pair was constructed by exploiting the Lie-Poisson structure on $so\left(3,n\right)$, created by the semi-direct sum of $so\left(3\right)$ and $n$ copies of $\mathbb{R}^{3}$.
% Such an approach would require a knowledge of the nonsemi-direct Lie algebra which corresponds to a generalised body force.
%
\par
It is interesting to note that Casimirs are `promoted' to first integrals (and thus become conditional on isotropy) as a new field is added in going to the next level of the hierarchy. For instance, at the second level of the hierarchy $\mathsf{n}$ is added as a uniform field and hence $\frac{1}{2}\mathsf{n}\cdot\mathsf{n}$ is a Casimir. In the next perturbation, by the field $\mathsf{B}$, the Casimir is perturbed to $\frac{1}{2}\mathsf{n}\cdot\mathsf{n}+\mathsf{m}\cdot\mathsf{B}$. After one more perturbation, by the field $\mathsf{D}$, this Casimir is turned into the first integral $\frac{1}{2}\mathsf{n}\cdot\mathsf{n}+\mathsf{m}\cdot\mathsf{B}+K\mathsf{D}\cdot\mathsf{d}_3$. By contrast, the Casimir $\mathsf{n}\cdot\mathsf{m}$ at the second level is perturbed directly into the integral $\mathsf{n}\cdot\mathsf{m}+K\mathsf{B}\cdot\mathsf{d}_3$
at the next level and remains the same one level up.
\par
We have not discussed boundary conditions of the equations. While rigid-body problems are initial-value problems, the rod problems considered here are typically two-point boundary-value problems. Buckling results for various types of boundary conditions can be found in Wolfe's work \cite{Wolfe88,Seidman88,Wolfe96} and in \cite{Heijden05}. It is natural to wonder how other perturbations such as extensibility and shearability affect the hierarchy of rod models described here. We intend to take up this question in future work.
\appendix
\section{The canonical transformation matrix} \label{app:transform}
Expressing the new canonical variables by inverting the non-canonical variables \eref{eq:canonical_magnetic}, \eref{eq:canonical_moment} and \eref{eq:canonical_force} yields
\begin{eqnarray}
\theta & = \cos^{-1} \frac{B_{3}^{}}{\sqrt{C_{3}}}, \nonumber \\
p_{\theta} & =  \frac{m_{1}B_{2}-m_{2}B_{1}}{\sqrt{C_{3}-B_{3}^{2}}}, \nonumber \\
\phi & = \tan^{-1}\frac{ -B_{2} }{ B_{1} }, \nonumber \\
p_{\phi} & = m_{3}, \nonumber \\
\psi & = \cos^{-1} \left( \frac{ n_{3}^{}-C_{2}^{}B_{3}^{}\slash C_{3} }{ \sqrt{\left(1-B_{3}^{2}\slash C_{3}^{} \right)\left(2C_{1}^{}-C_{2}^{2}\slash C_{3}^{}-2\left(m_{1}^{}B_{1}^{}+m_{2}^{}B_{2}^{}+m_{3}^{}B_{3}^{}\right)\right)} } \right), \nonumber \\
p_{\psi} & = \left( m_{1}B_{1}+m_{2}B_{2}+m_{3}B_{3}\right)\slash\sqrt{C_{3}}. \nonumber
\end{eqnarray} 
The transformation matrix $G$ in \eref{eq:G} is given by
\begin{eqnarray} 
\fl
G & = \left( 
\begin{array}{ccccccccc}
0 & 0 & 0 & 0 & 0 & 0 & 0 & 0 & -\sqrt{ \frac{C_{3}^{}}{1-C_{3}^{}B_{3}^{2}} } \\  
0 & 0 & 0 & 0 & 0 & 0 & \frac{B_{2}^{}}{B_{1}^{2}+B_{2}^{2}} & -\frac{B_{1}^{}}{B_{1}^{2}+B_{2}^{2}} & 0 \\ 
g_{31} & g_{32} & g_{33} & g_{34} & g_{35} & g_{36} & g_{37} & g_{38} & g_{39} \\ 
\frac{B_{2}^{}}{\sqrt{C_{3}-B_{3}^{}}} & -\frac{B_{1}}{\sqrt{C_{3}-B_{3}^{2}}} & 0 & 0 & 0 & 0 & -\frac{m_{2}}{\sqrt{C_{3}-B_{3}^{2}}} & \frac{m_{1}}{\sqrt{C_{3}-B_{3}^{2}}} & \frac{\left(m_{1}^{}B_{2}^{}-m_{2}B_{1}\right)B_{3}^{}}{\left(C_{3}-B_{3}^{2}\right)^{3\slash2}} \\ 
0 & 0 & 1 & 0 & 0 & 0 & 0 & 0 & 0 \\ 
B_{1}^{}\slash{\sqrt{C_{3}}} & B_{2}^{}\slash{\sqrt{C_{3}}} & B_{3}^{}\slash{\sqrt{C_{3}}} & 0 & 0 & 0 & m_{1}\slash{\sqrt{C_{3}}} & m_{2}\slash{\sqrt{C_{3}}} & m_{3}\slash{\sqrt{C_{3}}}
\end{array} 
\right), \nonumber
\end{eqnarray}
where
\begin{eqnarray}
g_{31} & = -\frac{ B_{1}\left( n_{3}-C_{2}B_{3}\slash{C_{3}} \right)}{ \Delta }, \nonumber \\
g_{32} & = -\frac{ B_{2}\left( n_{3}-C_{2}B_{3}\slash{C_{3}} \right)}{ \Delta }, \nonumber \\
g_{33} & = -\frac{ B_{3}\left( n_{3}-C_{2}B_{3}\slash{C_{3}} \right)}{ \Delta }, \nonumber \\
g_{34} & = 0, \nonumber \\
g_{35} & = 0, \nonumber \\
g_{36} & = -\frac{ 1 }{ \Delta }, \nonumber \\
g_{37} & = -\frac{ m_{1}\left( n_{3}-C_{2}B_{3}\slash{C_{3}} \right)}{ \Delta }, \nonumber \\
g_{38} & = -\frac{ m_{2}\left( n_{3}-C_{2}B_{3}\slash{C_{3}} \right)}{ \Delta }, \nonumber \\
g_{39} & = \frac{ C_{2}\left(2C_{1}^{}-C_{2}^{2}\slash{C_{3}} - 2\mathsf{m}\cdot\mathsf{B}\right)\slash{C_{3}} }{ \Delta }, \nonumber \\
& \hspace{1.0cm} - \frac{ \left(n_{3}-C_{2}B_{3}\slash{C_{3}}\right)\left( B_{3}\left(2C_{1}-C_{2}^{2}\slash{C_{3}}-2\mathsf{m}\cdot\mathsf{B}\right)-m_{3}\left(1-B_{3}\slash{C_{2}}\right) \right) }{ \Delta } . \nonumber 
\end{eqnarray}
with the denominator given by
\begin{eqnarray}
\fl
\Delta & = \left(2C_{1}^{}-C_{2}^{2}\slash{C_{3}} - 2\mathsf{m}\cdot\mathsf{B}\right) \sqrt{ \left(1-B^{2}_{3}\slash{C_{3}}\right)\left(2C_{1}^{}-C_{2}^{2}\slash{C_{3}}-2\mathsf{m}\cdot\mathsf{B} \right) - \left(n_{3}^{}-C_{2}^{}B_{3}^{}\slash{C_{3}} \right)^{2} }.\nonumber 
\end{eqnarray}
The non-canonical structure matrix is given by 
\begin{eqnarray}
J & = \left( 
\begin{array}{cccccccccc}
0 & -m_{3} & m_{2} & 0 & -n_{3} & n_{2} & 0 & -B_{3} & B_{2} \\
m_{3} & 0 & -m_{1} & n_{3} & 0 & -n_{1} & B_{3} & 0 & -B_{1} \\
-m_{2} & m_{1} & 0 & -n_{2} & n_{1} & 0 & -B_{2} & B_{1} & 0\\
0 & -n_{3} & n_{2} & 0 & - B_{3} & B_{2} & 0 & 0 & 0 \\
n_{3} & 0 & -n_{1} & B_{3} & 0 & -B_{1} & 0 & 0 & 0 \\
-n_{2} & n_{1} & 0 & -B_{2} & B_{1} & 0 & 0 & 0 & 0 \\
0 & -B_{3} & B_{2} & 0 & 0 & 0 & 0 & 0 & 0\\
B_{3} & 0 & -B_{1} & 0 & 0 & 0 & 0 & 0 & 0\\
-B_{2} & B_{1} & 0 & 0 & 0 & 0 & 0 & 0 & 0
\end{array}
\right), \nonumber
\end{eqnarray}
while the canonical structure matrix is given by
\begin{eqnarray}
\bar{J} & = \left( 
\begin{array}{ccccccc}
0 & 0 & 0 & 1 & 0 & 0 \\
0 & 0 & 0 & 0 & 1 & 0 \\
0 & 0 & 0 & 0 & 0 & 1 \\
-1 & 0 & 0 & 0 & 0 & 0 \\
0 & -1 & 0 & 0 & 0 & 0 \\
0 & 0 & -1 & 0 & 0 & 0 
\end{array}
\right). \nonumber
\end{eqnarray}
\section*{References}
\bibliographystyle{unsrt}
\bibliography{Bibliography}
\end{document}